# Molecular simulation study of the heat capacity of metastable water between 100K and 300K

J. Puibasset

*ICMN, UMR7374, CNRS, Université d'Orléans, Orléans, France ; 1b, rue de la Férollerie, 45071, Orléans cedex 2; puibasset@cnrs-orleans.fr*

P. Judeinstein

*Laboratoire Léon Brillouin, CEA, CNRS, Université Paris-Saclay ;CEA Saclay, 91191 Gif-sur-Yvette Cedex, France.*

J.-M. Zanotti

*Laboratoire Léon Brillouin, CEA, CNRS, Université Paris-Saclay ;CEA Saclay, 91191 Gif-sur-Yvette Cedex, France.*

# Molecular simulation study of the heat capacity of metastable water between 100K and 300K

Molecular simulations have been used to study the heat capacity of metastable liquid water at low temperature adsorbed on a smooth surface. These calculations aim at modelling water properties measured by experiments performed on water films adsorbed on Vycor nanoporous silica at low temperature. In particular, the study focuses on the non-monotonous variation of the heat capacity around between 100 and 300 K.

**Introduction**

In bulk water, local energetics and long-range connectivity come along with a very specific local three-dimensional (3D) tetrahedral organization of the hydrogen bond (HBond) network. Altogether these properties drive the numerous so-called "water anomalies",[1,2] for example, the apparent power-law divergence of the specific heat (Cp) at low temperature.

In interfacial situations, as the water molecules are forced to organize in only two dimensions, the three dimensional extension of the HBond network is frustrated. This profoundly affects the physics of water and a number of events are experimentally detected at 150, 220 and 250 K.[3] Thermal events are detected at these three temperatures in experimental Cp measurement,[4] but no clear-cut interpretation of the phenomenon has been reached so far.

Computer simulation of water is widely used to provide important information on the structure at the molecular level, as well as dynamical and thermodynamic properties. The quality of the intermolecular potential is of course mandatory for reliable results. Many water models have been developed, the simplest being rigid with fixed partial charges, to more sophisticated ones including flexible bonds and polarizability.[5,6] One of the most popular models is probably the SPC model.[7] It has been intensively used in molecular simulations involving large systems because it is a fast model in terms of computational cost. The heat capacity of SPC water has been shown to be in reasonable agreement with other simulation models as well as experiments.[8]

The water adsorption in nanoporous silica has been extensively studied by molecular simulations, in particular in the grand canonical ensemble, or the generalization of the Gibbs ensemble.[9-18] It is well-known that the fluid adsorption is influenced by the chemical disorder of the adsorbing surface.[19-23] In particular, the structure and dynamics of an adsorbed water film has been shown to be influenced by the nature of the silica surface.[15,24] This is expected to have an impact on the heat capacity of adsorbed layers of water, to be studied now.

**Numerical details**

*Model and interactions*

The system to be studied is a water film adsorbed on a plane surface. The interactions between water molecules are described by the SPC model.[7] This fast computable model is well suited as it reproduces quite well the thermodynamic properties of water around ambient temperature, like vapor pressure, enthalpy of vaporization and heat capacity.[25] In this model, a Lennard-Jones center is located on the oxygen atom, and positive and negative charges are placed on the three atoms. The water-water contribution to energy is calculated directly with minimal image convention, with a cutoff for the LJ contribution situated at 0.95 nm. Long range corrections are not applied for the LJ term. However, long range electrostatic contributions are handled with the Ewald summation. The direct space parameter kappa = 2.288 nm$^{-1}$ and $7^3$ reciprocal vectors are chosen. The total expected precision is around 1%. The simulation box is cubic for the bulk and the films configurations. The number of molecules for the bulk simulations is 216, with dimensions L = 1.863 nm, while for the film configurations there are 220 molecules on a surface with dimensions L=3.72 nm.

The film configurations correspond to water adsorbed on a hydrophilic surface, typically silica. The interaction between water molecules and silica species contain both van der Waals and electrostatic contributions due to hydroxyls at the silica surface. However, in this preliminary study we will neglect the atomic roughness of the adsorbing surface (smooth wall approximation), as well as the presence of hydroxyl groups. As a result, the external potential felt by the adsorbed water molecules is that produced by a continuum of silica atoms in a half-space interacting with the water oxygen species via the 12-6 LJ potential. The integration results in the 9-3 potential

$$V(z) = \frac{4}{3}\pi\varepsilon_{sf}\rho_s\sigma_{sf}^3\left[\frac{\sigma_{sf}^9}{15z^9} - \frac{\sigma_{sf}^3}{2z^3}\right] \quad (1)$$

where $\rho_s$ is the density of the solid interaction sites, and $\varepsilon_{sf}$ and $\sigma_{sf}$ are the parameters of the 12-6 LJ interaction between water molecules and silica.[9] However, the external potential obtained with this procedure is far too weak to be representative of a hydrophilic surface, because of the ignored surface dipoles associated with hydroxyls which are responsible for a major contribution to the interaction. As a consequence, we have chosen to enhance the external potential so that the minimum equals -15 kJ/mol, representative of a hydrophilic silica surface.

*Adsorption isotherm and heat capacity*

The adsorption isotherm is calculated as follows. The initial configuration is the empty simulation box. We set the chemical potential at its lowest value, corresponding to an external vapor pressure of one tenth of the saturating pressure $P_{sat}$ ($P/P_{sat} = 0.1$). The grand canonical Monte Carlo (GCMC) procedure is applied until the equilibrium is reached. After equilibration is reached, a new GCMC is started for a larger value of the chemical potential (corresponding to $P/P_{sat} = 0.2$), the initial configuration of the new run being the final configuration of the previous one. The same procedure is repeated until the saturating chemical potential is reached. These GCMC calculations are devised to produce and equilibrate the water film used to calculate the heat capacity.

The heat capacity is calculated in the canonical ensemble, *i.e.* for a given film thickness, or, equivalently, for a fixed number of molecules. Note however that the system contains a liquid/vapor interface, which is free to move to equilibrate pressure and chemical potential between the liquid film and the vapor. The heat capacity is obtained from the standard fluctuations formula of the intermolecular enthalpy during the Monte Carlo run.[26] The quantum corrections are taken into account so that the simulation results can be compared to experiments.[27] The simulations are started at the highest temperature, the final configuration of a given run being used as the initial configuration for a lower temperature simulation. This procedure is thought to be the most efficient to equilibrate the system at low temperature. Long Monte Carlo runs are however required to reach the equilibrium, in particular at low temperature where mobility of the molecules is extremely low. We have performed $10^6$ Monte Carlo steps per molecule for each temperature for equilibration, followed by $5 \times 10^6$ steps per molecule for data acquisition.

**Results and discussion**

*Adsorption isotherm at 300K*

The GCMC results for the adsorption isotherm are given in Figure 1. The amount of water adsorbed increases with the chemical potential, and exhibits a step around $P/P_{sat} = 0.3$. This corresponds to the formation of the first layer of water. A second step is visible around $P/P_{sat} = 0.8$, which is associated with the formation of a second layer of water molecules on the surface. This evolution of the amount adsorbed is in agreement with previous calculations done by taking into account explicitly the electrostatic contributions due to partial charges of the silica species, in particular the presence of hydroxyls.[24] The extremely simple model used in the present study to describe the surface thus catches the main features of a hydrophilic surface.

The heat capacity of the water film is calculated for two different film thicknesses corresponding approximately to 22 µmol/m$^2$ (monolayer, P/P$_{sat}$ = 0.5) and 44 µmol/m$^2$ (two layers, P/P$_{sat}$ = 0.8). Visual inspection of the molecular configurations (see Figure 2) shows that the first water layer is highly structured, with OH parallel to the adsorbing surface. As expected, the second layer is much less structured.

*Heat capacity*

The heat capacity dependence with temperature is given in Figure 3, for the two water film thicknesses adsorbed on the plane surface. As can be seen, the two curves exhibit non monotonic variations with the temperature. The general trend is that the heat capacity is independent of the film thickness at low and high temperature (below 150 K and above 300 K), while strong differences appear in the intermediate region. Both curves show an increase of the heat capacity, with a maximum around 225 K. However, the peak for the thin film is more pronounced, and probably narrower than for the thick film. How do these results compare with the bulk and experiments? The bulk simulation data show an almost monotonic increase of the heat capacity with temperature, except for a small excess between 200 and 250 K. Taking into account the rather large error bars of the simulations (10-20%) this excess might be the signature of an underlying peak around 225 K. Note that it is well known that the SPC water model predicts an increase of the heat capacity when reducing the temperature in a much lesser extent than experiments.[8] An interesting feature emerges when comparing the results for the bulk and the two film thicknesses: the narrower the film, the higher the peak. This suggests a strong correlation between the appearance of this peak and the structure of the first layer of adsorbed water. It is furthermore emphasized that the smooth surface used to model silica enhances the structuration of the first layer. These results are thus expected to be altered for a more realistic atomistic silica surface.

Another feature to be noticed is that at low temperature, the heat capacity of water converges towards the same value, independent of the film thickness, the value being that of the bulk water. This value remains significantly higher than the expected value for ice: the SPC water model is frozen in a metastable state even for the lowest temperature. The simulations are unable to cross the nucleation barrier associated with the crystallization of water.

An interesting point would be to check for the robustness of these results with the model. It is thus planned to perform similar calculations with other water models (TIP4P, TIP4P/2005) and another silica model including an atomistic description and silanols.

**References**

(1)     Debenedetti, P. G.; Stanley, H. E. Supercooled and Glassy Water. *Physics Today* **2003**, *56*, 40-46.

(2)     Caupin, F. Escaping the No Man's Land: Recent Experiments on Metastable Liquid Water. *J. Non-Cryst. Solids* **2015**, *407*, 441-448.

(3)     Zanotti, J.-M.; Judeinstein, P.; Dalla-Bernardina, S.; Creff, G.; Brubach, J.-B.; Roy, P.; Bonetti, M.; Ollivier, J.; Sakellariou, D.; Bellissent-Funel, M.-C. Competing Coexisting Phases in 2d Water. *Sci. Rep.* **2016**, *6*, 25938.

(4)     Tombari, E.; Ferrari, C.; Salvetti, G.; Johari, G. P. Dynamic and Apparent Specific Heats During Transformation of Water in Partly Filled Nanopores During Slow Cooling to 110 K and Heating. *Thermochim. Acta* **2009**, *492*, 37-44.

(5)     Guillot, B. A Reappraisal of What We Have Learnt During Three Decades of Computer Simulations on Water. *J. Mol. Liquids* **2002**, *101*, 219-260.

(6) Ouyang, J. F.; Bettens, R. P. A. Modelling Water: A Lifetime Enigma. *Chimia* **2015**, *69*, 104-111.

(7) Berendsen, H. J. C.; Postma, J. P. M.; van Gunsteren, W. F.; Hermans, J. In *Intermolecular Forces*; B. Pullman: Dordrecht: Reidel, 1981; pp 331.

(8) Jorgensen, W. L.; Jenson, C. Temperature Dependence of TIP3P, SPC, and TIP4P Water from NPT Monte Carlo Simulations: Seeking Temperatures of Maximum Density. *J. Comp. Chem.* **1998**, *19*, 1179-1186.

(9) Puibasset, J.; Pellenq, R. J.-M. Water Adsorption on Hydrophilic Mesoporous and Plane Silica Substrates: A Grand Canonical Monte Carlo Simulation Study. *J. Chem. Phys.* **2003**, *118*, 5613-5622.

(10) Puibasset, J.; Pellenq, R. J.-M. A Grand Canonical Monte Carlo Simulation Study of Water Adsorption on Vycor-Like Hydrophilic Mesoporous Silica at Different Temperatures. *J. Phys.: Condens. Matter* **2004**, *16*, S5329-S5343.

(11) Puibasset, J. Thermodynamic Characterization of Fluids Confined in Heterogeneous Pores by Monte Carlo Simulations in the Grand Canonical and the Isobaric-Isothermal Ensembles. *J. Phys. Chem. B* **2005**, *109*, 8185-8194.

(12) Puibasset, J. Phase Coexistence in Heterogeneous Porous Media : A New Extension to Gibbs Ensemble Monte Carlo Simulation Method. *J. Chem. Phys.* **2005**, *122*, 134710.

(13) Mischler, C.; Horbach, J.; Kob, W.; Binder, K. Water Adsorption on Amorphous Silica Surfaces: A Car-Parrinello Simulation Study. *J. Phys.: Condens. Matter* **2005**, *17*, 4005-4013.

(14) Malani, A.; Ayappa, K. G.; Murad, S. Influence of Hydrophilic Surface Specificity on the Structural Properties of Confined Water. *J. Phys. Chem. B* **2009**, *113*, 13825-13839.

(15) Argyris, D.; Cole, D. R.; Striolo, A. Dynamic Behavior of Interfacial Water at the Silica Surface. *J. Phys. Chem. C* **2009**, *113*, 19591-19600.

(16) Bonnaud, P. A.; Coasne, B.; Pellenq, R. J. M. Molecular Simulation of Water Confined in Nanoporous Silica. *J. Phys.: Condens. Matter* **2010**, *22*, 284110.

(17) Yamashita, K.; Daiguji, H. Molecular Simulations of Water Adsorbed on Mesoporous Silica Thin Films. *J. Phys. Chem. C* **2013**, *117*, 2084-2095.

(18) Yamashita, K.; Kashiwagi, K.; Agrawal, A.; Daiguji, H. Grand Canonical Monte Carlo and Molecular Dynamics Simulations of Capillary Condensation and Evaporation of Water in Hydrophilic Mesopores. *Mol. Phys.* **2017**, *115*, 328-342.

(19) Kuchta, B.; Llewellyn, P.; Denoyel, R.; Firlej, L. Monte Carlo Simulations of Krypton Adsorption in Nanopores: Influence of Pore-Wall Heterogeneity on the Adsorption Mechanism. *Low Temp. Phys.* **2003**, *29*, 880-882.

(20) Puibasset, J. Capillary Condensation in a Geometrically and a Chemically Heterogeneous Pore: A Molecular Simulation Study. *J. Phys. Chem. B* **2005**, *109*, 4700-4706.

(21) Puibasset, J. Influence of Surface Chemical Heterogeneities on Adsorption/Desorption Hysteresis and Coexistence Diagram of Metastable States within Cylindrical Pores. *J. Chem. Phys.* **2006**, *125*, 074707.

(22) Kuchta, B.; Firlej, L.; Marzec, M.; Boulet, P. Microscopic Mechanism of Adsorption in Cylindrical Nanopores with Heterogeneous Wall Structure. *Langmuir* **2008**, *24*, 4013-4019.

(23) Feng, Z.; Zhang, X.; Wang, W. Adsorption of Fluids in a Pore with Chemical Heterogeneities: The Cooperative Effect. *Phys. Rev. E* **2008**, *77*, 051603.

(24) Puibasset, J.; Pellenq, R. J.-M. A Comparison of Water Adsorption on Ordered and Disordered Silica Substrates. *Phys. Chem. Chem. Phys.* **2004**, *6*, 1933-1937.

(25)     Jorgensen, W. L.; Chandrasekhar, J.; Madura, J. D.; Impey, R. W.; Klein, M. L. Comparison of Simple Potential Functions for Simulating Liquid Water. *J. Chem. Phys.* **1983**, *79*, 926-935.

(26)     Jorgensen, W. L. Convergence of Monte Carlo Simulations of Liquid Water in the NPT Ensemble. *Chem. Phys. Lett.* **1982**, *92*, 405-410.

(27)     Owicki, J. C.; Scheraga, H. A. Monte Carlo Calculations in the Isothermal-Isobaric Ensemble. 1. Liquid Water. *J. Am. Chem. Soc.* **1977**, *99*, 7403-7412.

(28)     Angell, C. A.; Sichina, W. J.; Oguni, M. Heat Capacity of Water at Extremes of Supercooling and Superheating. *J. Phys. Chem.* **1982**, *86*, 998-1002.

(29)     Maruyama, S.; Wakabayashi, K.; Oguni, M. Thermal Properties of Supercooled Water Confined within Silica Gel Pores. *AIP Conf. Proc.* **2004**, *708*, 675-676.

Figure 1. Grand Canonical Monte Carlo results of SPC water adsorption isotherm (T = 300 K) on a smooth hydrophilic surface (eq. 1). The horizontal line at 22 µmol/m$^2$ correspond to a single water layer (P/P$_{sat}$ = 0.5) and at 44 µmol/m$^2$ to two water layers (P/P$_{sat}$ = 0.8).

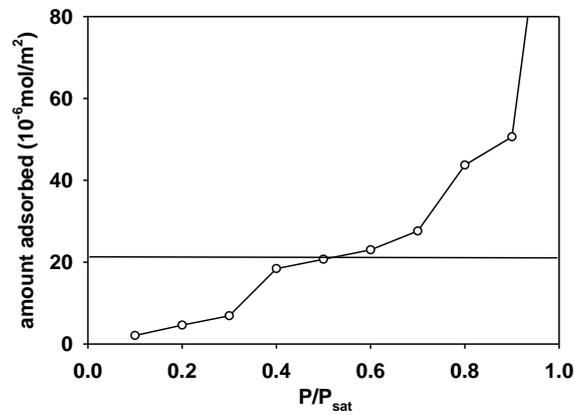

Figure 2. Molecular simulation configurations of SPC water at 300K adsorbed on the smooth hydrophilic surface (eq. 1). Left: thin film, corresponding to a single water layer (P/P$_{sat}$ = 0.5); Right: thick film, corresponding to two water layers (P/P$_{sat}$ = 0.8).

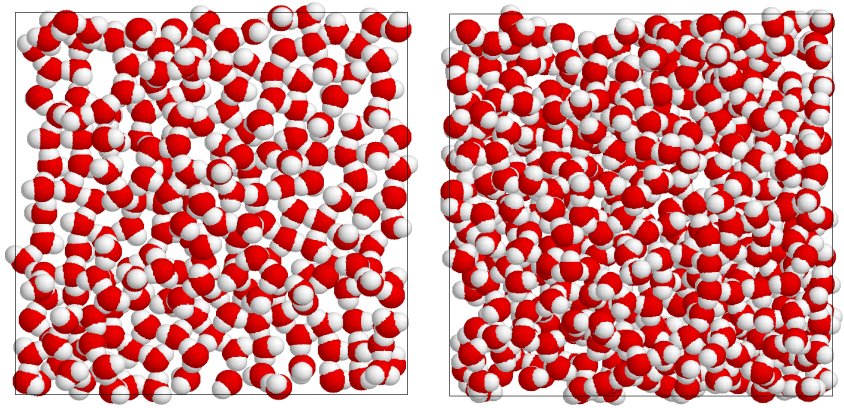

Figure 3. Heat capacity of the thin (green) and thick (blue) SPC water films adsorbed on the smooth hydrophilic surface. For comparison, the bulk (red) results are also given, as well as experimental data (black circles: metastable liquid water; up-triangles: ice).[28,29]

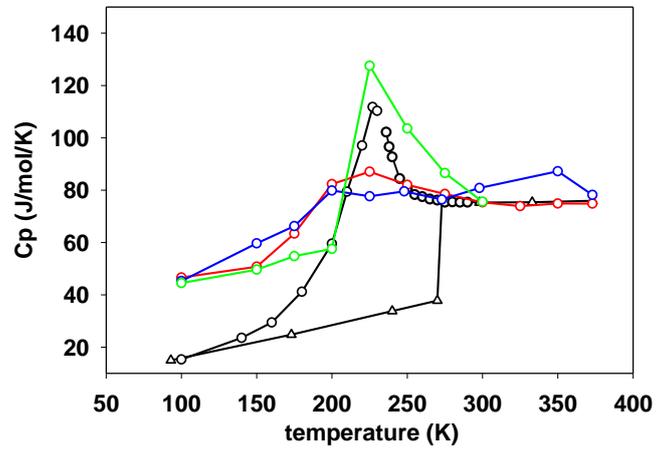